\documentstyle[prl,aps,multicol,epsfig]{revtex}

\newcommand{\boa}{{\bf a}}
\newcommand{\bq}{{\bf q}}
\newcommand{\bp}{{\bf p}}
\newcommand{\bz}{{\bf z}}
\newcommand{\ens}{{\cal E}}

\newcommand{\be}{\begin{equation}}

\newcommand{\ee}{\end{equation}}
\newcommand{\bs}{\begin{mathletters}} 
\newcommand{\es}{\end{mathletters}} 

\newcommand{\baa}{\begin{eqnarray}}
\newcommand{\eaa}{\end{eqnarray}}

\newcommand{\ba}{\bs\begin{eqnarray}}
\newcommand{\ea}{\end{eqnarray}\es}

\newcommand{\bt}[1]{\bs\label{#1}\begin{eqnarray}}
\newcommand{\et}{\end{eqnarray}\es}

\newcommand{\paper}[6]{#1 , #2 #3{\bf #4}, #5 (19#6)}

\newlength{\www}

\begin{document}

\title{Optimization of Trial Wave Functions for \\
Hamiltonian Lattice Models}
\author{Matteo Beccaria}
\address{
Dipartimento di Fisica, Universit\`a di Lecce, Via Arnesano,
73100 Lecce, Italy \\
Istituto Nazionale di Fisica Nucleare, Lecce
}
\maketitle
\begin{abstract}
We propose a new Monte Carlo algorithm for the numerical study of
general lattice models in Hamiltonian form. The algorithm is based on
an initial Ansatz for the ground state wave function depending on a
set of free parameters which are dynamically optimized. The method is
discussed in details  and results are
reported from explicit simulations of $U(1)$ lattice gauge theory in
$1+1$ dimensions.
\end{abstract}

\begin{multicols}{2}
\narrowtext

\section{Introduction}
\label{sec:intro}

The lattice formulation of Quantum Field Theory is a rigorous
theoretical framework for the study of non perturbative phenomena and 
its applications to phenomenologically relevant models, like QCD,   
are expected to play a major role in the next future.
In most numerical studies, the lattice version of a given model is
built in the Lagrangian formulation~\cite{Lagrangian}.
However, for certain ``static'' problems, like hadron
spectroscopy, the Hamiltonian approach~\cite{Hamiltonian} is more
suited because time is kept as continuous as possible and the
spectrum is deformed in a minimal way. 
Actually, even the most accurate Lagrangian calculations~\cite{Morningstar} 
evidence the advantages of using anisotropic lattices with smaller spacings in
the temporal direction, a typical feature of the Hamiltonian approach.

Another advantage of the Hamiltonian formulation 
is the possibility of easily exploiting any knowledge about 
the ground state wave function $\Psi_0$ in order to improve the
performance of numerical calculations. In particular, in this Letter, we are
concerned with the popular and all purpose Green Function Monte Carlo
algorithm (GFMC)~\cite{Linden} where an approximate version of
$\Psi_0$, the trial wave function $\Psi_T$, allows to speed up the 
algorithm convergence by means of Importance 
Sampling techniques~\cite{ImportanceSampling}. 

In a realistic problem, $\Psi_T$ depends 
on several free parameters $\boa=(a_1,\dots,a_N)$ that parametrize 
interaction terms responsible for correlations
expected to show up in the ground state.
The development of methods for the optimization of the free parameters $\boa$
is a key problem and an active field of research~\cite{Koch99}. 
The simplest approach is to perform a variational calculation
of the mean energy $E_V(\boa)~=~\langle \Psi_T(\boa) | H | \Psi_T(\boa)
\rangle$ where $H$ is the Hamiltonian. The quantity $E_V(\boa)$
is then minimized with respect to $\boa$. 
Since each evaluation of $E_V(\boa)$ requires a separate Monte Carlo
calculation, this is an expensive procedure for which
special tricks have been devised (e.g. correlated sampling~\cite{Ceperley77}).
A particularly interesting approach is that of~\cite{Harju97} where
the authors propose an algorithm for the 
automatic dynamical optimization of the free \underline{variational}
parameters $\boa$.

In this Letter, we abandon the variational approximation and consider
instead a
full Monte Carlo calculation for which we propose a new strategy to
optimize $\boa$. We describe an adaptive algorithm which converges on-line to
an optimal set of parameters $\boa^*$ which minimizes the statistical
error of the full (\underline{not} variational) Monte Carlo
simulation.
The non-adaptive core of the algorithm is a GFMC algorithm with
Importance Sampling and Stochastic Reconfiguration~\cite{SR}. 

Let us consider a quantum mechanical point particle with Hamiltonian
\be
H=H_0+V(\bq),\qquad H_0=\frac{1}{2} \bp^2
\ee
where $\bq=(q_1,\dots, q_d)$ is the position in ${\bf R}^d$ and 
$\bp=(p_1,\dots, p_d)$ is the associated momentum satisfying
the canonical commutation rules $[q_i,q_j]=0=[p_i,p_j]$ and
$[q_i,p_j]=i\delta_{ij}$. 
Let $\Psi_T(\bq, \boa)=\exp F(\bq, \boa)$ be a positive approximation
of the ground state wave function depending on some parameters
$\boa~=~(a_1, \dots, a_N)$. We perform a unitary transformation 
on $H$ based on $\Psi_T$ and build the new isospectral Hamiltonian 
\be
\widetilde H=\Psi_T H_0 \frac{1}{\Psi_T} + V(\bq) .
\ee
The Feynman propagator $\widetilde D=\langle \bq''|\exp(-t \widetilde
H)|\bq'\rangle$ 
associated with $\widetilde H$ can be written 
\baa
\lefteqn{\widetilde D(\bq'', \bq', \boa, \varepsilon) = 
\frac{1}{(2\pi\varepsilon)^{d/2}}\frac{1}{\sqrt{1+\varepsilon\nabla^2
F}}\times}&& \\
&\times& 
\exp\left\{
-(x_i[2\varepsilon(\delta_{ij}+\varepsilon\partial^2_{ij}
F)]^{-1} x_j-\varepsilon \widetilde V
\right\} + {\cal O}(\varepsilon^2), \nonumber \\
&& x_i=q''_i-q'_i-\varepsilon\partial_i F, \nonumber
\eaa
where 
\be
\widetilde V(\bq,\boa) = V(\bq)-\frac 1 2 (\nabla_\bq
F(\bq,\boa))^2-\frac 1 2 \nabla^2_\bq F(\bq,\boa)  .
\ee
By standard arguments, we conclude that $\widetilde D$
may be expressed in terms of a weighted average
\be
\label{GFMC}
\widetilde D(\bq'', \bq', \boa, t) =
\int_{\bq(0)=\bq', \bq(t)=\bq''}
\widetilde{\cal D} \bq(t)\ 
e^{-\int_0^t d\tau \widetilde V(\bq(\tau), \boa)} ,
\ee
where $\widetilde{\cal D}$ is the formal limiting measure (when
$\varepsilon\to 0$)
defined on the stochastic paths
generated according to the Langevin equation
\baa
\label{rwevol}
\bq_{n+1, i}&=&\bq_{n, i}+\varepsilon\partial_i F(\bq_n,\boa) + \\ 
&+& \sqrt{\varepsilon}(\bz_{n, i}+\frac 1 2 \varepsilon \partial^2_{ij}
F(\bq_n ,\boa) \bz_{n,j}) , \nonumber
\eaa
where $\bq_{n,i}$ denotes the i-th component of $\bq_n$ and 
$\bz_n$ are uncorrelated Gaussian random points in ${\bf R}^d$ with
unit variance.
The GFMC algorithm is expressed in a concise and formal way by
Eq.~(\ref{GFMC}), but
the actual calculation of the weighted averages must keep under
control the variance of the path-dependent weights
$\exp\{ -\int_0^t d\tau \widetilde V(\bq(\tau),\boa)\}$ which explodes as $t\to +\infty$.
An efficient technique to solve this problem is Stochastic
Reconfiguration~\cite{SR} which implements a kill and branch 
selection on the paths with the desirable feature of dealing always 
with a fixed size population of walkers. To this aim, 
a finite collection of $K$ walkers, an ensemble,  is introduced
\be
\ens = \{(\bq^{(n)}(t), \omega^{(n)}(t))\}_{1\le n\le K} ,
\ee
where the weights $\omega^{(n)}$ are defined by 
\be
\omega^{(n)}(t)=\exp -\int_0^t d\tau \widetilde V(\bq^{(n)}(\tau),\boa) .
\ee
The variance of the weights over the ensemble $\ens$ is 
\be
W(t)=\mbox{Var}\ \omega(t) = \frac{1}{K}\sum_{k=1}^K (\omega^{(k)}(t))^2 -
\left(\frac{1}{K}\sum_{k=1}^K \omega^{(k)}(t)\right)^2 ,
\ee
and the average of a function $f(\bq)$ over $\ens$ is 
\be
\langle f\rangle_\ens = \frac 1 K \sum_{k=1}^K f(\bq^{(k)})
\omega^{(k)} .
\ee
When $W(t)$ becomes too large, $\ens$ is transformed into a new
ensemble $\ens'$ with zero variance $W$ and the same averages,
at least in the $K\to\infty$ limit. Simulations can be
extended to arbitrarily large times with the drawback of a systematic
error vanishing with $K$ and an extrapolation to
$K\to\infty$ is required. More details on this procedure can be found in~\cite{SR}.

Statistical errors in such a Monte Carlo simulation are related to 
the fluctuations of $\widetilde V(\bq(\tau),\boa)$.
To see this, let us consider for instance the calculation of the ground state energy $E_0$. A simple algorithm
estimates the limit $E_0=\lim_{t\to +\infty} E(t,
\bq', \boa)$ where, for arbitrary $\bq'$ and $\boa$ , we define
\be
E(t, \bq',\boa) = -\frac{d}{dt} \log \int_{-\infty}^\infty \widetilde
D(\bq'', \bq', \boa, t) 
d\bq'' = \langle \widetilde V(\bq(t),\boa) \rangle ,
\ee
($\langle\cdot\rangle$ is the average over the weighted Langevin
trajectories as in Eq.~(\ref{GFMC})).
If $\Psi_T$ is an exact eigenstate of $H$ with eigenvalue $E$, then
we obtain $\widetilde
V\equiv E$ and the above formula gives $E(t) = E$
with zero variance, namely no statistical error.

In the more general case of non optimal $\Psi_T\neq \Psi_0$ a simulation 
performed with a sufficiently small $\varepsilon$ and a population of $K$
walkers will provide after $S$ Monte Carlo steps
only an approximate estimator $\hat E_0(S, K, \boa)$ of $E_0$, that is a random
variable with the asymptotic properties 
\baa
\langle \hat E_0(S, K, \boa) \rangle &=& E_0 + \frac{c_1(\boa)}{K^\alpha} +
o(K^{-\alpha}), \qquad \alpha>0, \\
\mbox{Var}\ \hat E_0(S, K, \boa) &=& \frac{c_2(K, \boa)}{\sqrt{S}},
\nonumber 
\eaa
where averages are over Monte Carlo realizations. 
The average of $\hat E_0$ extrapolated at $K\to\infty$ 
is exact and independent on
the trial parameters $\boa$ whereas $c_2(K,\boa)$ is in general 
strongly dependent on them and is expected to vanish when $K\to\infty$.

If the family  $\Psi_T(\bq, \boa)$ includes the exact ground state at the special point $\boa=\boa^*$, then
we know that $c_1(\boa^*)=c_2(K, \boa^*)=0$; in 
a less optimal situation, motivated by~\cite{Umrigar88}, we
seek a minimum of $c_2$. 

To establish an adaptive algorithm, we let $\boa \to \{\boa_n\}$ be a dynamical
parameter of the simulation and propose to update it together with the
ensemble according to the equations
\baa
\label{evol}
\bq_{n+1, i}&=&\bq_{n, i}+\varepsilon\partial_i F(\bq_n,\boa_n) +
\nonumber \\ 
&+& \sqrt{\varepsilon}(\bz_{n, i}+\frac 1 2 \varepsilon \partial^2_{ij}
F(\bq_n ,\boa_n) \bz_{n,j}) , \nonumber\\
\boa_{n+1} &=& \boa_n -\eta \nabla_\boa {\cal F}_n(\boa_n), \\
{\cal F}_n(\boa_n) &=& \mbox{Var}_\ens\ \widetilde V(\bq_n, \boa_n) , \nonumber
\eaa
where $\eta$ is a constant parameter and $\boa_n$ is the value of
$\boa$ at the n-th update.
In other words, we implement
a local minimization of the weight variance as a driving mechanism for
the free parameters.

The coupled set of equations Eqs.~(\ref{evol}) for the 
evolution of $\boa$ and the
random walkers is non linear and discrete. The optimization of $\boa$
and the GFMC are thus linked together.
This procedure can be successfully checked in trivial
quantum mechanical examples. Here we discuss a non trivial application
to test convergence and stability. The model we study is 
$U(1)$ lattice gauge theory in $1+1$ dimensions. Following the 
notation of~\cite{Barnes87} (see also~\cite{gauge} for other
applications of GFMC to lattice gauge theory), 
the Hamiltonian of the model is 
\be
H = \sum_{p=1}^L\left[ 
\left(\sum_{l_p=1}^3 -\frac{1}{2\beta}\frac{\partial^2}{\partial \theta_{l_p}^2}
\right) + \beta (1-\cos\phi_p)
\right],
\ee
where $\beta$ is the coupling constant, 
$L$ is the spatial lattice size, $\theta_{l_i,p}$ are link phases around the $p$-th plaquette and 
$\phi_p = \theta_{1, p}+\theta_{2, p+1}-\theta_{3, p}-\theta_{2, p}$
is the plaquette gauge invariant angle.
An accurate variational estimate of the ground state energy per
plaquette $E_0/L$ is obtained with the gauge invariant Ansatz 
$\Psi_0(\phi_1, \dots, \phi_L) = \exp\left(\lambda\sum_{p=1}^L
\cos\phi_p\right)$ leading to 
\be
\frac{E_0^{(var)}(\beta)}{L} = \beta + \frac 1 \beta \min_\lambda \left[(\lambda-\beta^2) 
\frac{I_1(2\lambda)}{I_0(2\lambda)}\right],
\ee
where $I_n$ is the n-th modified Bessel function.

We choose a rather general gauge invariant trial wave function $\Psi_T =
\exp F$ with inter plaquette correlations of the following form
\baa
F &=& \sum_{p=1}^L\{a_1\cos \phi_p + a_2\cos 2\phi_p + \\
&+& \sum_{k=1}^2 (a_{3,k} \cos (\phi_p+\phi_{p+k}) + a_{4,k}\cos
(\phi_p-\phi_{p+k})) \}, \nonumber
\eaa
which depends on 6 free parameters $a_1$, $a_2$, $a_{3,1}$, $a_{3,2}$,
$a_{4,1}$, $a_{4,2}$.
The particular case $a_{3,2}=a_{4,2}=0$ with only 
on-site and nearest-neighbour terms is discussed
in~\cite{Barnes87} where the ${\cal O}(\beta^4)$
perturbative optimal values are given:
$a_1^{pert} = \beta^2/2$, $a_2^{pert} = -\beta^4/32$, $a_{3,1}^{pert} = \beta^4/24$
and $a_{4,1}^{pert} = - \beta^4/40$; the ground state energy is measured by using
only $a_1$ which is optimized by trials and errors.  
In this Letter, we shall discuss the non perturbative behaviour of the
full six parameters at the equilibrium point reached automatically by
the algorithm.

We perform simulations with ensembles of $K=10$, $30$ and $50$ walkers on a
system with spatial lattice size $L=8$ and time step
$\varepsilon=0.015$. 
In the Lagrangian formulation this would correspond to a simulation on
a lattice with a very
large extension in the temporal direction.
The coupling $\beta$ is varied through the values
$0.5$, $1.0$, $1.5$, $2.0$ and $2.5$. We execute stochastic
reconfiguration of the ensemble each $r(\beta)$ temporal steps with
$r(\beta)$  ranging between a maximum 40 at $\beta=0.5$ and a minimum 20 at $\beta=2.5$.  
The learning parameter is $\eta=0.0005$ and all the free parameters
are zero at the first iteration. The number of Monte Carlo iterations is around
$5\times 10^4$ depending slightly on $\beta$.

In Fig.~(1) we show the Monte Carlo time history of the 
free parameters $\boa$ for the run with $\beta=0.5$ and $K=30$. After
about 300 iterations, the parameters reach an optimum value around
which they oscillate with small fluctuations.
If these fluctuations were too noisy, they 
could be removed by Stochastic Gradient Approximation
techniques, namely by letting $\eta$ be a time
dependent positive sequence $\{\eta_n\}_{n\ge 0}$ 
vanishing with $n\to\infty$ under the constraints 
$\sum_n \eta=\infty$ and 
$\sum_n \eta^2<\infty$~\cite{Harju97}.
The corresponding energy time history is shown in Fig.~(2) where
the fluctuations of the local measurements are shown to decrease very
rapidly.

The optimal values of the six parameters as functions of $\beta$ at
the largest $K$ used are shown in Tab.~(\ref{tavolaI}).
The leading perturbative expansion of $\boa$ 
can be checked to be correct only at the smallest $\beta$
and overestimates $\boa^*$ at larger $\beta$.
The next-to-nearest neighbour terms are rather small being at
$\beta=2.5$ roughly 1\% of the nearest neighbour ones.
The qualitative picture discussed in~\cite{Barnes87} is confirmed with
dominating plaquette anticorrelation in the ground state and
next-to-nearest neighbour effects below the percent level.
Here, we stress again, the determination of the optimal
set of parameters $\boa^*$ is completely automatic.

About the ground state energy, we show in Tab.~(\ref{tavolaII}) the
estimates computed with three values of $K$ to show the
very small systematic residual error associated with the finite size walker
population. We also show the Monte Carlo results from~\cite{Barnes87}
with which we agree within errors 
as well as the variational bound.
It is remarkable that a good estimate of the ground state energy
is obtained with the very small number of walkers $K=10$. This can 
be intepreted as a signal that for this admittedly simple model the
proposed six parameters wave function is rather accurate.

To conclude, let us remark that the above method is rather general and
is applicable to dynamically optimize the free parameters of a many body trial wave
function for any model that can be studied by GFMC with Stochastic
Reconfiguration. This includes for instance pure gauge $SU(2)$ and $SU(3)$ lattice
gauge theory in any dimension.

\begin{figure}
\caption{Monte Carlo time history of the 
free parameters $\boa$ for the run with $\beta=0.5$ and $K=30$.}
\epsfig{file=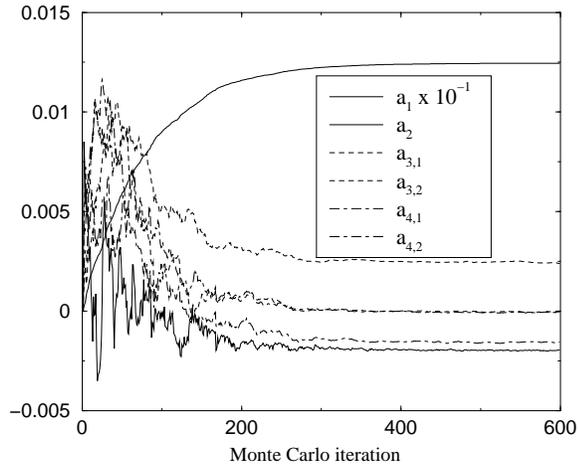,width=6.3cm,angle=-90}
\end{figure}\noindent

\begin{figure}
\caption{Monte Carlo time history of the energy measurements
during the first steps of the run with $\beta=0.5$ and $K=30$.}
\epsfig{file=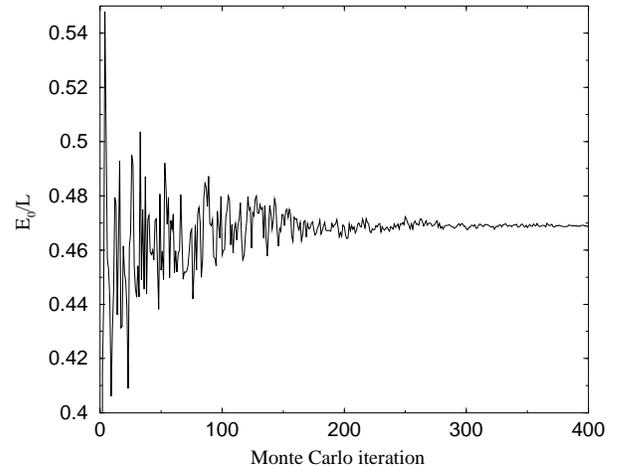,width=6.3cm,angle=-90}
\end{figure}\noindent

\end{multicols}
\widetext

\begin{table}
\caption{Optimal value of the 6 free parameters $a_1, a_2, a_{3,1},
a_{3,2}, a_{4,1}, a_{4,2}$ and
corresponding minimum standard deviation $\sigma_E$ of the ground
state energy estimator. We show the numerical results obtained with $50$ walkers. }
\label{tavolaI}
\begin{center}
\begin{tabular}{c|ccccccc}
$\beta$ & $a_1$ & $a_2$ &  $a_{3,1}$ & $a_{3,2}$ & $a_{4,1}$ &
$a_{4,2}$ & $\sigma_E$ \\
\hline
0.5&	0.1245271(7)&	-0.0019590(2)&	0.0024544(4)&	-0.0000374(3)&	-0.0015535(3)&	-0.0000609(3)&	0.0002\\
1.0&	0.46287(4)&	-0.022487(6)&	0.02597(1)&	0.00074(1)&	-0.015262(8)&	-0.00108(1)&	0.003\\
1.5&	0.8327(2)&	-0.05406(3)&	0.06255(5)&	0.00319(4)&	-0.02999(4)&	-0.00414(4)&	0.009\\
2.0&	1.1541(6)&	-0.07794(5)&	0.0997(1)&	0.00618(7)&	-0.03921(7)&	-0.00866(6)&	0.008\\
2.5&	1.4447(5)&	-0.09663(7)&	0.1370(1)&	0.01065(9)&	-0.0459(1)&	-0.0128(1)&	0.008\\
\hline  
\end{tabular}
\end{center}
\end{table}

\begin{table}
\caption{Comparison among the different estimates (the proposed
method, standard Monte Carlo and
variational) of the ground state energy. }
\label{tavolaII}
\begin{center}
\begin{tabular}{c|ccccc}
$\beta$ & $E_{\rm MC, K=10}^{\rm Adaptive}$ & $E_{\rm MC, K=30}^{\rm Adaptive}$ &
$E_{\rm MC, K=50}^{\rm Adaptive}$ & $E_{\rm MC, Barnes\ et\ al.}$ &
$E_{\rm Variational}$\\
\hline
0.5&	0.468930(2)&	0.468932(1)	&0.4689327(8)&	0.4690(1)&	0.4690\\
1.0& 	0.77032(3)&	0.77020(2)	&0.77024(2)&	0.7697(2)&	0.7746\\
1.5& 	0.88226(8)&	0.88209(5)	&0.88188(4)&	0.8823(8)&	0.9005\\
2.0& 	0.91718(9)&	0.91694(5)	&0.91705(4)&	0.916(1)&	0.9435\\
2.5& 	0.93245(8)&	0.93242(5)	&0.93230(3)&	0.929(2)&	0.9594\\
\hline  
\end{tabular}
\end{center}
\end{table}

\end{document}